\documentclass{sig-alternate-05-2015}
\usepackage{multirow}
\usepackage[british]{babel}

\begin{document}

\CopyrightYear{2016}
\setcopyright{acmcopyright}
\conferenceinfo{CIKM'16 ,}{October 24-28, 2016, Indianapolis, IN, USA}
\isbn{978-1-4503-4073-1/16/10}\acmPrice{\$15.00}
\doi{http://dx.doi.org/10.1145/2983323.2983769}

\clubpenalty=10000
\widowpenalty = 10000

\title{A Deep Relevance Matching Model for Ad-hoc Retrieval}

\author{
\alignauthor
Jiafeng Guo$^{\dagger}$,\hspace*{0.4cm}Yixing Fan$^{\dagger}$,\hspace*{0.4cm}Qingyao Ai$^{\ddagger}$,\hspace*{0.4cm}W. Bruce Croft$^{\ddagger}$\\
       \affaddr{$^{\dagger}$CAS Key Lab of Network Data Science and Technology, Institute of Computing Technology,}\\
       \affaddr{Chinese Academy of Sciences, Beijing, China}\\
       \affaddr{$^{\ddagger}$Center for Intelligent Information Retrieval, University of Massachusetts Amherst, MA, USA}\\
       \email{guojiafeng@ict.ac.cn, fanyixing@software.ict.ac.cn, $\{$aiqy,croft$\}$@cs.umass.edu}
}

\maketitle
\begin{abstract}
In recent years, deep neural networks have led to exciting breakthroughs in speech recognition, computer vision, and natural language processing (NLP) tasks. However, there have been few positive results of deep models on ad-hoc retrieval tasks. This is partially due to the fact that many important characteristics of the ad-hoc retrieval task have not been well addressed in deep models yet. Typically, the ad-hoc retrieval task is formalized as a matching problem between two pieces of text in existing work using deep models, and treated equivalent to many NLP tasks such as paraphrase identification, question answering and automatic conversation. However, we argue that the ad-hoc retrieval task is mainly about relevance matching while most NLP matching tasks concern semantic matching, and there are some fundamental differences between these two matching tasks. Successful relevance matching requires proper handling of the exact matching signals, query term importance, and diverse matching requirements. In this paper, we propose a novel deep relevance matching model (DRMM) for ad-hoc retrieval. Specifically, our model employs a joint deep architecture at the query term level for relevance matching. By using matching histogram mapping, a feed forward matching network, and a term gating network, we can effectively deal with the three relevance matching factors mentioned above. Experimental results on two representative benchmark collections show that our model can significantly outperform some well-known retrieval models as well as state-of-the-art deep matching models.
\end{abstract}

\keywords{Relevance Matching, Semantic Matching, Neural Models, Ad-hoc Retrieval, Ranking Models}

\section{Introduction}
Machine learning methods have been successfully applied to information retrieval (IR) in recent years. Typically, a ranking function which produces a relevance score given a query and document pair is learned based on a set of human defined features. However, handcrafting features can be time-consuming, incomplete and over-specified. On the other hand, deep neural networks, as a representation learning method, are able to discover from the training data the hidden structures and features at different levels of abstraction that are useful for the tasks. Recently, deep models have been applied to a variety of applications in computer vision \cite{lecun1995convolutional}, speech recognition \cite{hinton2012deep} and NLP \cite{socher2011dynamic,lu2013deep}, and have yielded significant performance improvements. Given the success of deep learning in these domains, it seems that deep learning should have a major impact on IR. However, there have been few positive results of deep models on IR tasks, especially ad-hoc retrieval tasks, until now.

Without loss of generality, when applying deep models to ad-hoc retrieval, the task is typically formalized as a matching problem between two pieces of text (i.e., the query and document). Such a matching problem formalization is often considered general in the sense that it can cover both ad-hoc retrieval tasks as well as many NLP tasks such as paraphrase identification, question answering (QA), and automatic conversation \cite{lu2013deep,hu2014convolutional}. A variety of deep matching models have been proposed to solve this matching problem, which can be categorized into two types according to their model architecture. One is the representation-focused model, which tries to build a good representation for a single text with a deep neural network, and then conducts matching between the compositional and abstract text representations. Examples include DSSM \cite{huang2013learning}, C-DSSM \cite{shen2014learning,gao2015modeling} and ARC-I \cite{hu2014convolutional}. The other is the interaction-focused model, which first builds local interactions (i.e., local matching signals) between two pieces of text, and then uses deep neural networks to learn hierarchical interaction patterns for matching. Examples include DeepMatch \cite{lu2013deep}, ARC-II \cite{hu2014convolutional} and MatchPyramid \cite{pang2016text}.

\begin{figure*}[tbp]
\centering
\includegraphics[scale=0.63]{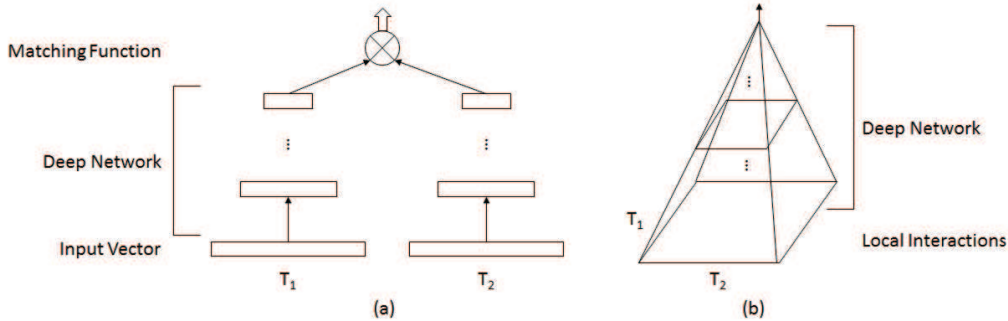}
\caption{Two types of deep matching models: (a) Representation-focused models employ a Siamese (symmetric) architecture over the text inputs; (b) Interaction-focused models employ a hierarchical deep architecture over the local interaction matrix.}\label{fig1}
\end{figure*}

However, in this work, we argue that the matching problems in many NLP tasks and the ad-hoc retrieval task are fundamentally different. Most NLP tasks concern semantic matching, i.e., identifying the semantic meaning and inferring the semantic relations between two pieces of text, while the ad-hoc retrieval task is mainly about relevance matching, i.e., identifying whether a document is relevant to a given query. We point out three major differences between these two matching problems which may lead to significantly different architecture design for the deep matching models. We also show that most existing deep matching models are designed for semantic matching rather than relevance \mbox{matching}.

Based on these differences, we propose a deep relevance matching model (DRMM) for ad-hoc retrieval by explicitly modeling the three major factors in relevance matching. Overall, our model is an interaction-focused model which employs a joint deep architecture at the query term\footnote{Here we use term to denote the indexed units in search systems, which could be stemmed words or phrases.} level for relevance matching. Specifically, we first build local interactions between each pair of terms from a query and a document based on term embeddings. For each query term, we map the variable-length local interactions into a fixed-length matching histogram. Based on this fixed-length matching histogram, we then employ a feed forward matching network to learn hierarchical matching patterns and produce a matching score. Finally, the overall matching score is generated by aggregating the scores from each query term with a term gating network computing the aggregation weights. We show how our major model designs, including matching histogram mapping, a feed forward matching network, and a term gating network, address the three key factors in relevance matching for ad-hoc retrieval.

We evaluate the effectiveness of the proposed DRMM based on two representative ad-hoc retrieval benchmark collections. For comparison, we take into account some well-known traditional retrieval models, as well as several state-of-the-art deep matching models either designed for the general matching problem or proposed specifically for the ad-hoc retrieval task. The empirical results show that the existing deep matching models cannot compete with the traditional retrieval models on these benchmark collections, while our model can outperform all the baseline models significantly in terms of all the evaluation metrics.

The major contributions of this paper include:
\begin{description}
\item{1.} We point out three major differences between semantic matching and relevance matching, which may lead to significantly different architecture design of the deep matching models.
\item{2.} We propose a novel deep relevance matching model for ad-hoc retrieval by explicitly addressing the three key factors of relevance matching.
\item{3.} We conduct rigorous comparisons over state-of-the-art retrieval models on benchmark collections and analyze the deficiencies of existing deep matching models and advantages of the DRMM.
\end{description}


\section{Ad-hoc Retrieval As A Matching Problem}
According to existing literature \cite{huang2013learning,lu2013deep}, the core problem in ad-hoc retrieval, i.e., the computation of the relevance for a document given a particular query, can be formalized as a text matching problem as follows. Given two texts $T_1$ and $T_2$, the degree of matching is typically measured as a score produced by a scoring function based on the representation of each text:
\begin{displaymath}
match(T_1,T_2)=F(\Phi(T_1),\Phi(T_2)),
\end{displaymath}
where $\Phi$ is a function to map each text to a representation vector, and $F$ is the scoring function based on the interactions between them. Such a text matching problem is considered general since it also describes many NLP tasks, such as paraphrase identification, question answering, and automatic conversation \cite{lu2013deep,hu2014convolutional}. A variety of deep matching models have been proposed either for the specific ad-hoc retrieval task or for the general matching problem.

Depending on how you choose the two functions, existing deep matching models can be categorized into two types. The first one, the representation-focused model, tries to build a good representation for a single text with a deep neural network, and then conducts matching between two compositional and abstract text representations. In this approach, $\Phi$ is a complex representation mapping function while $F$ is a relatively simple matching function. For example, in DSSM \cite{huang2013learning}, $\Phi$ is a feed forward neural network, while $F$ is the cosine similarity function. In C-DSSM \cite{shen2014learning,gao2015modeling}, $\Phi$ is a convolutional neural network (CNN) \cite{lecun1995convolutional}, while $F$ is the cosine similarity function. In ARC-I \cite{hu2014convolutional}, $\Phi$ is a CNN, while $F$ is a multi-layer perceptron (MLP). Without loss of generality, all the model architectures of representation-focused models can be viewed as a Siamese (symmetric) architecture over the text inputs, as shown in Figure \ref{fig1}(a).

The second one, the interaction-focused model, first builds the local interactions between two texts based on some basic representations, and then uses deep neural networks to learn the hierarchical interaction patterns for matching. In this approach, $\Phi$ is usually a simple mapping function while $F$ is a complex deep model. For example, in DeepMatch \cite{lu2013deep}, $\Phi$ simply maps each text to a sequence of words, while $F$ is a feed forward neural network powered by a topic model over the word interaction matrix. In ARC-II \cite{hu2014convolutional} and MatchPyramid \cite{pang2016text}, $\Phi$ maps each text to a sequence of word vectors, while $F$ is a CNN over the interaction matrix between word vectors from the two texts. Without loss of generality, all the model architectures of interaction-focused models can be viewed as a hierarchical deep architecture over the local interaction matrix, as shown in Figure \ref{fig1}(b).

Although various deep matching models have been proposed under such a general matching problem formalization, most of them have only been demonstrated to be effective on a set of NLP tasks such as paraphrase identification and QA \cite{hu2014convolutional,wan2015deep}. There have been few positive results on the ad-hoc retrieval task. Even the deep models specially designed for Web search, e.g., DSSM and C-DSSM, were only evaluated on <query, doc title> pairs which are not a typical ad-hoc retrieval setting. If we directly apply these deep matching models on some benchmark retrieval collections, e.g. TREC collections, we find relatively poor performance compared to traditional ranking models, such as the language model \cite{zhai2001study} and BM25 \cite{robertson1994some}. All these observations raise some questions such as: Is matching in ad-hoc retrieval really the same as that in NLP tasks? Are the existing deep matching models suitable for the ad-hoc retrieval task?

\section{Semantic Matching vs. Relevance Matching}
In this section, we discuss the differences between text matching in ad-hoc retrieval and other NLP tasks. The matching in many NLP tasks, such as paraphrase identification, question answering and automatic conversation, is mainly concerned with \textit{semantic matching}, i.e., identifying the semantic meaning and inferring the semantic relations between two pieces of text. In these semantic matching tasks, the two texts are usually homogeneous and consist of a few natural language sentences, such as questions/answer sentences, or dialogs. To infer the semantic relations between natural language sentences, semantic matching emphasizes the following three factors:

\textbf{Similarity matching signals:} It is important, or critical to capture the semantic similarity/relatedness between words, phrases and sentences, as compared with exact matching signals. For example, in paraphrase identification, one needs to identify whether two sentences convey the same meaning with different expressions. In automatic conversation, one aims to find a proper response semantically related to the previous dialog, which may not share any common words or phrases between them.

\textbf{Compositional meanings:} Since texts in semantic matching usually consist of natural language sentences with grammatical structures, it is more beneficial to use the compositional meaning of the sentences based on such grammatical structures rather than treating them as a set/sequence of words \cite{socher2011dynamic}. For example, in question answering, most questions have clear grammatical structures which can help identify the compositional meaning that reflects what the question is about.

\textbf{Global matching requirement:} Semantic matching usually treats the two pieces of text as a whole to infer the semantic relations between them, leading to a global matching requirement. This is partially related to the fact that most texts in semantic matching have limited lengths and thus the topic scope is concentrated. For example, two sentences are considered as paraphrases if the whole meaning is the same, and a good answer fully answers the question.

The matching in ad-hoc retrieval, on the contrary, is mainly about \textit{relevance matching}, i.e., identifying whether a document is relevant to a given query. In this task, the query is typically short and keyword based, while the document can vary considerably in length, from tens of words to thousands or even tens of thousands of words. To estimate the relevance between a query and a document, relevance matching is focused on the following three factors:

\textbf{Exact matching signals:} Although term mismatch is a critical problem in ad-hoc retrieval and has been tackled using different semantic similarity signals, the exact matching of terms in documents with those in queries is still the most important signal in ad-hoc retrieval due to the indexing and search paradigm in modern search engines. For example, Fang and Zhai \cite{fang2006semantic} proposed the semantic term matching constraint which states that matching an original query term exactly should always contribute no less to the relevance score than matching a semantically related term multiple times.
This also explains why some traditional retrieval models, e.g., BM25, can work reasonably well purely based on exact matching signals.

\textbf{Query term importance:} Since queries are mainly short and keyword based without complex grammatical structures in ad-hoc retrieval, it is important to take into account term importance, while the compositional relation among the query terms is usually the simple ``and'' relation in operational search. For example, given the query ``bitcoin news'', a relevant document is expected to be about ``bitcoin'' and ``news'', where the term ``bitcoin'' is more important than ``news'' in the sense that a document describing other aspects of ``bitcoin'' would be more relevant than a document describing ``news'' of other things. In the literature, there have been many formal studies on retrieval models showing the importance of term discrimination \cite{fang2004formal,fang2011diagnostic}.

\begin{figure*}[tbp]
\centering
\includegraphics[scale=0.6]{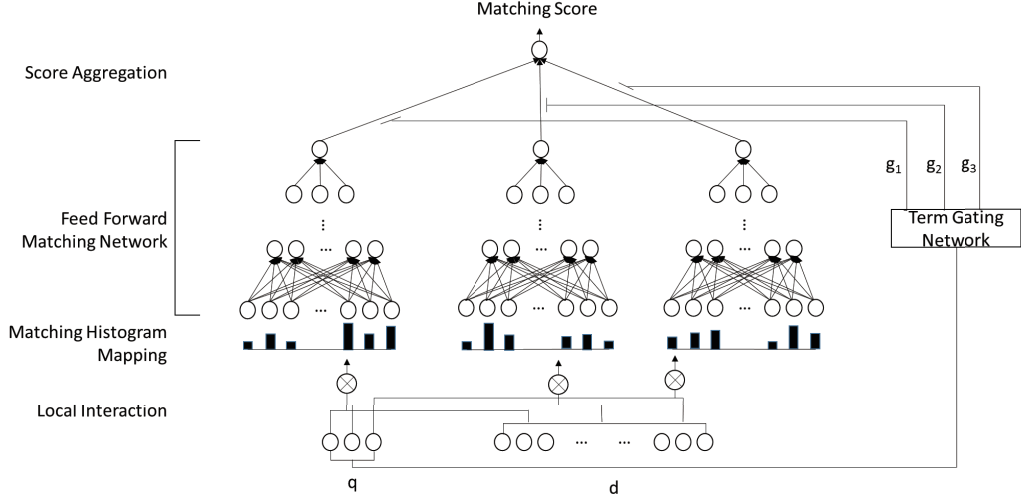}
\caption{Architecture of the Deep Relevance Matching Model.}\label{fig2}
\end{figure*}

\textbf{Diverse matching requirement:} In ad-hoc retrieval, a relevant document can be very long and there have been different hypotheses concerning document length \cite{robertson1994some} in the literature, leading to a diverse matching requirement. Specifically, the \textit{Verbosity Hypothesis} assumes that a long document is like a short document, covering a similar scope but with more words. In this case, the relevance matching might be global if we assume short documents have a concentrated topic. On the contrary, the \textit{Scope Hypothesis} assumes a long document consists of a number of unrelated short documents concatenated together. In this way, the relevance matching could happen in any part of a relevant document, and we do not require the document as a whole to be relevant to a query.

As we can see, there are significant differences between relevance matching in ad-hoc retrieval and semantic matching in many NLP tasks. These differences affect the design of deep model architectures and it may be difficult to find a ``one-fit-all'' solution to such different matching problems. If we revisit the existing deep matching models, we find that most of them concern semantic matching rather than relevance matching. For example, the representation-focused models such as DSSM, C-DSSM and ARC-I focus on the compositional meaning of the texts and fit the global matching requirement. In these models, detailed matching signals and, especially, exact matching signals are lost since they defer the interaction between two texts until their individual representations have been created \cite{hu2014convolutional}. Although the interaction-focused models such as DeepMatch, ARC-II and MatchPyramid preserve both exact and similarity matching signals, they do not differentiate these signals but treat them as equally important. These models focus on learning the composition of local interactions without addressing term importance. In particular, the convolutional structures in ARC-II and MatchPyramid are designed to learn positional regularities, which may work well under the global matching requirement but fail under the diverse matching requirement.(There is more discussion on this in Section 4.)

\section{Deep Relevance Matching Model}
Based on the above analysis, we propose a novel deep matching model specifically designed for relevance matching in ad-hoc retrieval by explicitly addressing the three factors described in Section 3. We refer to our model as a deep relevance matching model (DRMM). Overall, our model is similar to interaction-focused models rather than representation-focused models since the latter would inevitably lose the detailed matching signals which are critical for relevance matching in ad-hoc retrieval.

Specifically, our model employs a joint deep architecture at the query term level over the local interactions between query and document terms for relevance matching. We first build local interactions between each pair of terms from a query and a document based on term embeddings. For each query term, we then transform the variable-length local interactions into a fixed-length matching histogram. Based on the fixed-length matching histogram, we employ a feed forward matching network to learn hierarchical matching patterns and produce a matching score for each query term. Finally, the overall matching score is generated by aggregating the scores from each single query term with a term gating network computing the aggregation weights. The model architecture is depicted in Figure \ref{fig2}.

More formally, suppose both query and document are represented as a set of term vectors denoted by $q\!\!=\!\!\{w^{(q)}_1,\dots,w^{(q)}_M\}$ and $d=\{w^{(d)}_1,\dots,w^{(d)}_N\}$, where $w^{(q)}_i, i=1,\dots,M$ and $w^{(d)}_j, j=1,\dots,N$ denotes a query term vector and a document term vector, respectively, and $s$ denotes the final relevance score, we have
\begin{eqnarray}
\boldsymbol{z}_i^{(0)} = h(w^{(q)}_i\otimes d),&&\!\!\!\!\quad i\!\!=1,\dots,M\nonumber\\
\boldsymbol{z}_i^{(l)} = tanh(\boldsymbol{W}^{(l)}\boldsymbol{z}_i^{(l-1)}+\boldsymbol{b}^{(l)}),&&\!\!\!\!\quad i\!\!=1,\dots,M, l\!\!=1,\dots,L\nonumber\\
s = \sum_{i=1}^{M}g_iz_i^{(L)}&&\nonumber
\end{eqnarray}
where $\otimes$ denotes the interaction operator between a query term and the document terms, $h$ denotes the mapping function from local interactions to matching histogram, $\boldsymbol{z}_i^{(l)}, l=0,\dots,L$ denotes the intermediate hidden layers for the $i$-th query term, and $g_i, i=1,\dots,M$ denotes the aggregation weight produced by the term gating network. $\boldsymbol{W}^{(l)}$ denotes the $l$-th weight matrix and $\boldsymbol{b}^{(l)}$ denotes the $l$-th bias term, which are shared across different query terms. Note that we adopt cosine similarity, a widely used measure for semantic closeness in neural embeddings \cite{mikolov2013distributed,pennington2014glove}, as the interaction operator between each pair of term vectors from a query and a document. In our work, we assume the term vectors are learned a priori using existing neural embedding models such as Word2Vec \cite{mikolov2013distributed}. We do not learn term vectors in our deep relevance matching model for the following reasons: 1) Reliable term representations can be better acquired from large scale unlabeled text collections rather than from the limited ground truth data for ad-hoc retrieval; 2) By using the a priori learned term vectors, we can focus the learning of our model on relevance matching patterns and considerably reduce the model complexity.
In the following, we will describe the major components of our model, including the matching histogram mapping, feed forward matching network, and term gating network in detail, and discuss how they address the three key factors of relevance matching in ad-hoc retrieval.

\textbf{Matching Histogram Mapping:} The input of our deep relevance matching model is the local interactions between each pair of terms from a query and a document. A major problem is that the size of local interactions is not fixed due to the varied lengths of queries and documents.
Previous interaction-based models view the local interactions as a matching matrix by preserving the sequential term orders in both queries and documents. Clearly the matching matrix is a position preserving representation, which is useful if the learning task is position related. However, according to the diverse matching requirement, relevance matching is not position related since it could happen in any position in a long document. Thus the matching matrix may not be a suitable representation for ad-hoc retrieval due to the potentially noisy positional signals in it.

In our work, we adopt a strength preserving representation, namely a matching histogram, which groups local interactions according to different levels of signal strengths rather than their positions. Specifically, since the local interaction (i.e., cosine similarity between term vectors) is within the interval $[-1,1]$, we discretize the interval into a set of ordered bins and accumulate the count of local interactions in each bin. In this work, we consider fixed bin size and treat exact matching as a separate bin. Other discretization schemes could be explored in future work. For example, suppose the bin size is set as $0.5$, we will obtain five bins $\{[-1,-0.5), [-0.5,-0), [0,0.5),$ $[0.5,1), [1,1]\}$ in an ascending order. Given a query term ``\textit{car}'' and a document $(car, rent, truck, bump, injunction, runway)$, and the corresponding local interactions based on cosine similarity are $(1, 0.2, 0.7, 0.3, -0.1, 0.1)$, we will obtain a matching histogram as $[0, 1, 3, 1, 1]$.
We explore three ways of the matching histogram mapping:
\begin{description}
\item[Count-based Histogram (CH):] This is the simplest way of transformation as described above which directly takes the count of local interactions in each bin as the histogram value.
\item[Normalized Histogram (NH):] We normalize the count value in each bin by the total count to focus on the relative rather than the absolute number of different levels of interactions.
\item[LogCount-based Histogram (LCH):] We apply logarithm over the count value in each bin, both to reduce the range, and to allow our model to more easily learn multiplicative relationships \cite{burges2005learning}.
\end{description}

We compare our matching histogram representation with previous matching matrix representations to show the advantages. Firstly, by setting exact matching as a separate bin, the matching histogram clearly distinguishes the exact matching signals from similarity matching signals, while in a matching matrix all the signals are mixed together. Secondly, to solve the problem of variable size in the matching matrix, a zero-padding scheme is often adopted in previous methods \cite{hu2014convolutional}. However, the zero-padding scheme introduces additional interaction signals which may be unfair for short documents. In contrast, we map the variable-size interactions into a fixed-length matching histogram without introducing any additional signals. 

\textbf{Feed forward Matching Network:} Based on the matching histogram above, we employ a feed forward matching network to learn the hierarchical matching patterns and produce a matching score for each query term. Since our model follows the approach of interaction-focused models, we discuss the major differences between the learning of our feed forward matching network and that in previous interaction-focused models.

Existing interaction-focused models, e.g., ARC-II and MatchPyramid, employ a CNN to learn hierarchical matching patterns over the matching matrix. These models are basically position-aware using convolutional units with a local ``receptive field'' and learning positional regularities in matching patterns. This may be suitable for the image recognition task, and work well on semantic matching problems due to the global matching requirement (i.e., all the positions are important). However, it may not be suitable for the ad-hoc retrieval task, since such positional regularity may not exist in relevance matching due to the diverse matching requirement discussed in Section 3. Besides, since CNN parameters are position related, these models will treat both exact matching and similarity matching signals equally.

Our deep relevance matching model, on the contrary, aims to extract hierarchical matching patterns from different levels of interaction signals rather than different positions. The position-free and strength-focused property makes it better at handling the diverse matching requirement in ad-hoc retrieval. Meanwhile, since the matching histogram directly distinguishes exact matching signals from the rest, our model can naturally learn the importance of exact matching signals.

There have been some interaction-focused models that employ special pooling strategies to turn the position-aware interactions into strength-based fixed-length representations. For example, MV-LSTM \cite{wan2015deep} used $K$-max pooling strategy \cite{kalchbrenner2014convolutional} to select the top $K$ strongest interaction signals from the matching matrix as the input of a MLP. However, such a pooling strategy simply truncates the signals and thus will be strongly biased to long documents since it is more likely for long documents to contain more strong signals. The pooling strategy is applied over the entire matching matrix in MV-LSTM, making it possible that the top $K$ strongest signals all come from the interactions between a single query term and the document terms. In contrast, our model does not rely on any pooling strategy to truncate the interactions so that we can avoid these problems.

\textbf{Term Gating Network:} One significant difference of our model from existing interaction-focused models is that we employ a joint deep architecture at the query term level. In this way, our model can explicitly model query term importance. This is achieved by using the term gating network, which produces an aggregation weight for each query term controlling how much the relevance score on that query term contributes to the final relevance score. Specifically, we employ the softmax function as the gating function.
\begin{displaymath}
g_i=\frac{\exp(\boldsymbol{w}_g\boldsymbol{x}^{(q)}_i)}{\sum_{j=1}^{M}\exp(\boldsymbol{w}_g\boldsymbol{x}^{(q)}_j)}, \quad i=1,\dots,M,
\end{displaymath}
where $\boldsymbol{w}_g$ denotes the weight vector of the term gating network and $\boldsymbol{x}^{(q)}_i, i=1,\dots,M$ denotes the $i$-th query term input. We tried different inputs for the gating function as follows:

\textbf{Term Vector (TV):} Inspired by the work \cite{zheng2015learning} where term embeddings can be leveraged to learn the term weights in queries, we use query term vectors as the input of the gating function. In this method, $\boldsymbol{x}^{(q)}_i$ denotes the $i$-th query term vector, and $\boldsymbol{w}_g$ is a weight vector with the same dimensionality of term vectors.

\textbf{Inverse Document Frequency (IDF):} An important signal of term importance in ad-hoc retrieval is the inverse document frequency. We also tried this simple but powerful signal in the gating function. In this method, $\boldsymbol{x}^{(q)}_i$ denotes the inverse document frequency of the $i$-th query term, and $\boldsymbol{w}_g$ reduces to a single parameter.

\subsection{Model Training}
Since the ad-hoc retrieval task is fundamentally a ranking problem, we employ a pairwise ranking loss such as hinge loss to train our deep relevance matching model. Given a triple $(q,d^+,d^-)$, where document $d^+$ is ranked higher than document $d^-$ with respect to query $q$, the loss function is defined as:
\begin{equation*}
\mathcal{L}(q,d^+,d^-;\Theta) = \max(0, 1-s(q, d^+)+s(q, d^-))
\end{equation*}
where $s(q,d)$ denotes the predicted matching score for $(q,d)$, and $\Theta$ includes the parameters for the feed forward matching network and those for the term gating network. The optimization is relatively straightforward with standard back-propagation \cite{williams1986learning}. We apply stochastic gradient descent method Adagrad \cite{duchi2011adaptive} with mini-batches ($20$ in size), which can be easily parallelized on single machine with multi-cores. For regularization, we find that the early stopping \cite{giles2001overfitting} strategy works well for our model. 

\section{Experiments}
In this section, we conduct experiments to demonstrate the effectiveness of our proposed model. 

\subsection{Data Sets}
To conduct experiments, we use two TREC collections, Robust04 and ClueWeb-09-Cat-B.
The details of the two collections are provided in Table \ref{tab:data_stats}. As we can see, they represent different sizes and genres of heterogeneous text collections. Robust04 is a small news dataset. Its topics are collected from TREC Robust Track 2004. ClueWeb-09-Cat-B, on the other hand, is a large Web collection, whose topics are accumulated from TREC Web Tracks 2009, 2010, and 2011. Note that ClueWeb-09-Cat-B is filtered to the set of documents with spam scores in the $60^{th}$ percentile, using the Waterloo Fusion spam scores \cite{cormack2011efficient}. For both datasets, we made use of both the title and the description of each TREC topic in our experiments. The retrieval experiments described in this section are implemented using the Galago Search Engine\footnote{http://www.lemurproject.org/galago.php}. During indexing and retrieval, both documents and query words are white-space tokenized, lowercased, and stemmed using the Krovetz stemmer \cite{krovetz1993viewing}. Stopword removal is performed on query words during retrieval using the INQUERY stop list \cite{callan1995trec}.

\begin{table}
  \centering
  \caption{Statistics of the TREC collections used in this study. The ClueWeb-09-Cat-B collection has been filtered to the set of documents in the $60^{th}$ percentile of spam scores.}
  \begin{tabular}{c c c c}
  \hline
    & Robust04 &   ClueWeb-09-Cat-B \\\hline
    Vocabulary & 0.6M & 38M  \\
    Document Count & 0.5M   & 34M  \\
    Collection Length & 252M  & 26B  \\
    Query Count & 250 & 150\\\hline
  \end{tabular}
  \label{tab:data_stats}
\end{table}

\subsection{Baselines and Experimental Settings}
We adopt three types of baseline methods for comparison, including traditional retrieval models, representation-focused deep matching models and interaction-focused deep matching models. Traditional retrieval models include

\textbf{QL:} Query likelihood model based on Dirichlet smoothing \cite{zhai2001study} is one of the best performing language models. 

\textbf{BM25:} The BM25 formula \cite{robertson1994some} is another highly effective retrieval model that represents the classical probabilistic retrieval model.

Representation-focused deep matching models include

\textbf{DSSM$_T$/DSSM$_D$:} DSSM \cite{huang2013learning} is a state-of-the-art deep matching model for Web search. In the original paper, the model was evaluated based on <query, doc title> pairs where doc title is extracted from the title field. We denote this model as DSSM$_T$. Since other baseline models and our model are based on the full text of the documents, we also evaluated the DSSM model under the same setting, denoted by DSSM$_D$. Since DSSM needs large scale training data due to its huge parameter size, we directly used the released model\footnote{http://research.microsoft.com/en-us/downloads/731572aa-98e4-4c50-b99d-ae3f0c9562b9/} (trained on large click-through dataset) in our experiments.

\textbf{C-DSSM$_T$/C-DSSM$_D$:} C-DSSM \cite{shen2014learning,gao2015modeling} is a similar deep matching model to DSSM for Web search, replacing the feed forward neural network with a convolutional neural network. For the same reason as DSSM, we also made use of the released model$^{3}$ directly and adopt two versions of the C-DSSM model, one based on title fields of documents denoted as C-DSSM$_T$ and the other based the whole document denoted as C-DSSM$_D$. 

\textbf{ARC-I:} ARC-I \cite{hu2014convolutional} is a general representation-focused deep matching model that has been tested on a set of NLP tasks including sentence completion, response matching, and paraphrase identification. We implemented the ARC-I model according to the original paper since there is no publicly available code.

Interaction-focused deep matching models are as follows:

\textbf{ARC-II:} ARC-II \cite{hu2014convolutional} was proposed by the authors of the model ARC-I, but focuses on learning hierarchical matching patterns from local interactions using a CNN. We also implemented ACR-II since there is no publicly available code.

\textbf{MP:} MatchPyramid \cite{pang2016text} is another state-of-the-art inter-action-focused deep matching model and has been tested on two NLP tasks including paraphrase identification and paper citation matching. There are three variants of the model based on different interaction operators, denoted as MP$_{IND}$, MP$_{COS}$, and MP$_{DOT}$. We obtained the original implementation of the model from the authors for comparison. 

We refer to our proposed deep relevance matching model as \textbf{DRMM}. With different types of histogram mapping functions (i.e., CH, NH and LCH) and term gating functions (i.e., TV and IDF), we obtained six different variants of our proposed model. For example, by DRMM$_{CH\times IDF}$ we refer to DRMM with Count-based histogram and term gating network using inverse document frequency.

\textbf{Term Embeddings:} For all the models based on term embedding inputs, including ARC-I, ARC-II, MatchPyramid and DRMM, we used $300$-dimensional term vectors trained with the Continuous Bag-of-Words (CBOW) Model \cite{mikolov2013distributed} on the Robust04 and ClueWeb-09-Cat-B collections, respectively. Specifically, we used $10$ as the context window size and used $10$ negative samples and a subsampling of frequent words with sampling threshold of $10^{-4}$ as suggested by Word2Vec\footnote{https://code.google.com/p/word2vec/}. Each corpus was pre-processed by removing HTML tags and stemming. We also discarded from the vocabulary all the terms that occur less than $10$ times in the corpus, which resulted in a vocabulary of size $0.1$M and $4.1$M on the Robust04 and ClueWeb-09-Cat-B collections, respectively. To address the out-of-vocabulary (OOV) terms (i.e., some rare terms or numbers not trained by CBOW) in queries, we follow the practice in previous work \cite{kenter2015short} to only allow exact matching between such query terms and document terms.

\begin{table*}[!ht]\centering
  \caption{Comparison of different retrieval models over the Robust-04 and ClueWeb-09-Cat-B collections. Significant improvement or degradation with respect to QL is indicated (+/-) ($p$-$value \le 0.05$).}
  \begin{tabular}{c c l l l l l l l}
  \multicolumn{9}{c}{Robust-04 collection} \\
  \hline \hline
    & & \multicolumn{3}{c}{Topic titles} & & \multicolumn{3}{c}{Topic descriptions} \\\cline{3-5}\cline{7-9}
    Model Type & Model Name & MAP & nDCG@20 & P@20  & & MAP & nDCG@20 & P@20 \\\hline
    \multirow{2}{3cm}{\centering Traditional Retrieval Baselines} & QL & 0.253 & 0.415 & 0.369  & & 0.246 & 0.391 & 0.334 \\
    & BM25 & 0.255 & 0.418 & 0.370 & & 0.241 & 0.399 & 0.337  \\\hline
    \multirow{3}{3.4cm}{\centering Representation-Focused Matching Baselines}& DSSM$_D$ & 0.095$^-$  & 0.201$^-$ & 0.171$^-$ &  &  0.078$^-$ & 0.169$^-$ & 0.145$^-$ \\
    & CDSSM$_D$ & 0.067$^-$ & 0.146$^-$ & 0.125$^-$ &  & 0.050$^-$ & 0.113$^-$ & 0.093$^-$  \\
    & ARC-I & 0.041$^-$ & 0.066$^-$ & 0.065$^-$  &  &  0.030$^-$  & 0.047$^-$ &  0.045$^-$  \\\hline
    \multirow{4}{3cm}{\centering Interaction-Focused Matching Baselines}& ARC-II & 0.067$^-$ & 0.147$^-$ & 0.128$^-$ &  & 0.042$^-$ & 0.086$^-$ & 0.074$^-$ \\
    & MP$_{IND}$ & 0.169$^-$ & 0.319$^-$ & 0.281$^-$ & & 0.067$^-$ & 0.142$^-$ & 0.118$^-$ \\
    & MP$_{COS}$ & 0.189$^-$ & 0.330$^-$ & 0.290$^-$ &  &0.094$^-$ & 0.190$^-$ & 0.162$^-$  \\
    & MP$_{DOT}$ & 0.083$^-$ & 0.159$^-$ & 0.155$^-$ & & 0.047$^-$ & 0.104$^-$ & 0.092$^-$  \\\hline
    \multirow{6}{3cm}{\centering Our Approach} & DRMM$_{CH\times TV}$ & 0.253 & 0.407 & 0.357 & & 0.247 & 0.404 & 0.341 \\
    &DRMM$_{NH\times TV}$ & 0.160$^-$ & 0.293$^-$ & 0.258$^-$ & & 0.132$^-$  & 0.217$^-$ & 0.186$^-$ \\
    &DRMM$_{LCH\times TV}$ & 0.268$^+$ & 0.423 & 0.381 & & 0.265$^+$ & 0.423$^+$ & 0.360$^+$ \\
    &DRMM$_{CH\times IDF}$ & 0.259 & 0.412 & 0.362 & & 0.255 & 0.410$^+$ & 0.344 \\
    &DRMM$_{NH\times IDF}$ & 0.187$^-$ & 0.312$^-$ & 0.282$^-$ & & 0.145$^-$ & 0.243$^-$ & 0.199$^-$ \\
    &DRMM$_{LCH\times IDF}$ & \textbf{0.279}$^+$ & \textbf{0.431}$^+$ & \textbf{0.382}$^+$ & & \textbf{0.275}$^+$ & \textbf{0.437}$^+$ & \textbf{0.371}$^+$  \\\hline\hline\\
     \multicolumn{9}{c}{ClueWeb-09-Cat-B collection} \\
  \hline \hline
    & & \multicolumn{3}{c}{Topic titles} & & \multicolumn{3}{c}{Topic descriptions} \\\cline{3-5}\cline{7-9}
    Model Type & Model Name & MAP & nDCG@20 & P@20  & & MAP & nDCG@20 & P@20 \\\hline
    \multirow{2}{3cm}{\centering Traditional Retrieval Baselines}& QL & 0.100 & 0.224 & 0.328  & & 0.075 & 0.183 & 0.234 \\
    & BM25 & 0.101 & 0.225 & 0.326 & & 0.080 & 0.196 & 0.255$^+$  \\\hline
    \multirow{5}{3.4cm}{\centering Representation-Focused Matching Baselines}& DSSM$_T$ & 0.054$^-$ & 0.132$^-$ & 0.185$^-$ &  &0.046$^-$ & 0.119$^-$ & 0.143$^-$  \\
    & DSSM$_D$ & 0.039$^-$ & 0.099$^-$ & 0.131$^-$ & & 0.034$^-$ & 0.078$^-$ & 0.103$^-$ \\
    & CDSSM$_T$ & 0.064$^-$ & 0.153$^-$ & 0.214$^-$ & & 0.055$^-$ & 0.139$^-$ & 0.171$^-$ \\
    & CDSSM$_D$ & 0.054$^-$ & 0.134$^-$ & 0.177$^-$ & & 0.049$^-$ & 0.125$^-$ & 0.160$^-$ \\
    & ARC-I & 0.024$^-$ & 0.073$^-$ & 0.089$^-$ & & 0.017$^-$ & 0.036$^-$ & 0.051$^-$\\\hline
    \multirow{4}{3cm}{\centering Interaction-Focused Matching Baselines}& ACR-II & 0.033$^-$ & 0.087$^-$ & 0.123$^-$ &  & 0.024$^-$ & 0.056$^-$ & 0.075$^-$ \\
    & MP$_{IND}$ &  0.056$^-$ & 0.139$^-$ & 0.208$^-$ & & 0.043$^-$ & 0.118$^-$ & 0.158$^-$ \\
    & MP$_{COS}$ & 0.066$^-$ & 0.158$^-$ & 0.222$^-$ & & 0.057$^-$ & 0.140$^-$ & 0.171$^-$ \\
    & MP$_{DOT}$ & 0.044$^-$ & 0.109$^-$ & 0.158$^-$ & & 0.033$^-$ & 0.073$^-$ & 0.102$^-$ \\\hline
    \multirow{6}{3cm}{\centering Our Approach} & DRMM$_{CH\times TV}$ & 0.103 & 0.245 & 0.347 & & 0.072 & 0.188 & 0.253\\
    &DRMM$_{NH\times TV}$ & 0.065$^-$ & 0.151$^-$ & 0.213$^-$ & & 0.031$^-$ & 0.075$^-$ & 0.100$^-$\\
    &DRMM$_{LCH\times TV}$ & 0.111$^+$ & 0.250$^+$ & 0.361$^+$ & & 0.083 & 0.213 & 0.275  \\
    &DRMM$_{CH\times IDF}$ & 0.104 & 0.252$^+$ & 0.354$^+$ & & 0.077 & 0.204 & 0.267 \\
    &DRMM$_{NH\times IDF}$ & 0.066$^-$ & 0.151$^-$ & 0.216$^-$ & & 0.038$^-$ & 0.087$^-$ & 0.113$^-$ \\
    &DRMM$_{LCH\times IDF}$ & \textbf{0.113}$^+$ & \textbf{0.258}$^+$ & \textbf{0.365}$^+$ & & \textbf{0.087}$^+$ & \textbf{0.235}$^+$ & \textbf{0.310}$^+$ \\\hline\hline\\
  \end{tabular}
  \label{tab:rank_result}
\end{table*}

\textbf{Network Configurations:} For network configurations (e.g., numbers of layers and hidden nodes), we tune the hyper parameters on a validation set (as part of the training set). For ARC-I, ARC-II and MatchPyramid, we tried both the default configurations in their original paper and other settings. We find that models with less layers and feature maps perform better, probably due to the limited training data in TREC collections. Specifically, for ARC-I and ARC-II, we use $3$-word windows, $64$ feature maps and $6$ layers (two for convolutions, two for max-pooling and two full connection). For MatchPyramid, we use one convolutional layer, one dynamic pooling layer and two full connection layers. The number of feature maps is $8$ and the kernel size is set to be $3\times 3$. For DRMM, we also use a four-layer architecture throughout all experiments, i.e., one histogram input layer ($30$ nodes), two hidden layers in the feed forward matching network ($5$ nodes and $1$ node respectively), and one output layer ($1$ node) with the term gating network for the final matching score.

\subsection{Evaluation Methodology}
Given the limited number of queries for each collection, we conduct $5$-fold cross-validation to minimize over-fitting without reducing the number of learning instances. Topics for each collection are randomly divided into $5$ folds. The parameters for each model are tuned on $4$-of-$5$ folds. The final fold in each case is used to evaluate the optimal parameters. This process is repeated $5$ times, once for each fold. Mean average precision (MAP) is the optimized metric for all retrieval models. Throughout this paper each displayed evaluation statistic is the average of the five fold-level evaluation values. For evaluation, the top-ranked $1,000$ documents are compared using the mean average precision (MAP), normalized discounted cumulative gain at rank $20$ (nDCG@20), and precision at rank $20$ (P@20). Statistical differences between models are computed using the Fisher randomization test~\cite{smucker2007comparison} ($\alpha=0.05$). Note that for all the deep matching models, we adopt a re-ranking strategy for efficient computation. An initial retrieval is performed using the QL model to obtain the top $2,000$ ranked documents. We then use the deep matching models to re-rank these top results. The top-ranked $1,000$ documents are then used for comparison.

\subsection{Retrieval Performance and Analysis}
This section presents the performance results of \mbox{different} retrieval models over the two benchmark datasets. A summary of results is displayed in Table \ref{tab:rank_result}.

As we can see, all the representation-focused models perform significantly worse than the traditional retrieval models, demonstrating the unsuitability of these models for relevance matching. Both DSSM$_T$ and C-DSSM$_T$ can work better than their counterpart on the whole document on ClueWeb-09-Cat-B, showing that models designed for global matching requirement cannot handle the diverse matching requirement in long documents. Note that we do not report the performance of DSSM$_T$ and C-DSSM$_T$ on Robust04 since there is no title field in many subsets in this collection. The ARC-I model, although trained on the corresponding corpus, performs even worse than DSSM and C-DSSM. A possible reason is that ARC-I concatenates the query and document representation for computing the matching score, which may be less effective than the cosine function in DSSM and C-DSSM.

When we look at the interaction-focused models, we find that these baseline models cannot compete with the traditional retrieval models either. Among these models, ARC-II can outperform ARC-I by directly learning from local interactions, but performs worse than MatchPyramid models due to the indirect local interactions (i.e., local interaction is based on the weighted sum of query and document term vectors rather than cosine similarity or dot product), which is consistent with previous results in \cite{hu2014convolutional,pang2016text}. Moreover, the best performing interaction-focused model, MP$_{COS}$, can consistently outperform all the representation-focused models on both test collections. When comparing the MatchPyramid models, we find that both MP$_{IND}$ and MP$_{COS}$ perform much better than MP$_{DOT}$. Note that MP$_{IND}$ is purely based on exact matching signals, MP$_{COS}$ and MP$_{DOT}$ involve both exact and similarity matching signals where exact matching signals are always stronger than similarity signals in MP$_{COS}$, but this may not be true in MP$_{DOT}$. The performance gap between MP$_{DOT}$ and the other two MPs indicates the importance of the exact matching signals in relevance matching. In fact, when evaluated on the semantic matching tasks in \cite{pang2016text}, MP$_{DOT}$ performed better than the other two MPs even though it cannot differentiate the exact matching signals from the rest, demonstrating the significant differences between semantic matching and relevance matching.

As for our proposed DRMMs, we have the following observations: (1) NH-based models perform significantly worse than CH-based models, while LCH-based models achieve the best performance on both collections. The low performance of NH-based models may be related to the loss of document length information after normalization which is important in ad-hoc retrieval \cite{fang2011diagnostic}. Meanwhile, the good performance of LCH-based models indicates that deep neural networks can benefit from input signals with reduced range and non-linear transformation useful for learning multiplicative relationships \cite{burges2005learning}; (2) The term gating function based on inverse document frequency works better than that based on term vectors. There are two possible reasons for this result. Firstly, term vectors do not contain sufficient information for the term importance. Secondly, the learning of the model might be dominated by the term gating network when we use term vectors as the input since there are more parameters (i.e., $300$ parameters) in the gating network compared to the feed forward matching network (i.e., $155$ parameters). 

Finally, we can see that the best performing DRMM (i.e., DRMM$_{LCH\times IDF}$) is significantly better than all the existing deep matching models as well as traditional retrieval \mbox{models}. For example, on ClueWeb-09-Cat-B topic titles, the relative improvement of our model over the best performing baseline (i.e., BM25) is about $11.9\%$, $14.7\%$, and $12\%$ in terms of MAP, nDCG@20 and P@20, respectively. Another interesting finding is that on the Robust04 collection, the performance of DRMM$_{LCH\times IDF}$ on topic descriptions can be comparable to that on topic titles, which is seldom observed on previous models. This also demonstrates the potential of our model in handling long queries in ad-hoc retrieval.

\subsection{Analysis on DRMM model}
We conducted experiments to verify the effectiveness of different components in the DRMM and analyze the effect of term embedding dimensions. Through these experiments, we try to gain a better understanding of the DRRM.

\subsubsection{Impact of Different Model Components}
To study the effect of different model components, we compare the original DRMM$_{LCH\times IDF}$ with several simpler versions of the model. Firstly, we removed the term gating network and used a simple sum to aggregate the scores from all the query terms. Since the aggregation weight is uniform, we denote this model as DRMM$_{LCH\times UNI}$. We also tried removing the histogram mapping layer but kept the rest unchanged. To turn the variable-length local interactions into a fixed-length representation, we adopted two pooling strategies. One is dynamic pooling as in \cite{socher2011dynamic,pang2016text} which keeps the position information, and the other is $K$-max pooling as in \cite{wan2015deep} which turns the positional signals into strength related signals. For a fair comparison, we require the size of the representation after pooling to be the same as the size of the matching histogram (i.e., $30$). Note that although the matching network structure is the same, the learned model is significantly different due to the change of the input. The matching model based on dynamic pooling is a position-aware model, while the model based on $K$-max pooling is learned with respect to the top strong interaction signals. We denote the former model as DRMM$_{DYN\times IDF}$ and the latter as DRMM$_{KMAX\times IDF}$.
\begin{figure}[tbp]
\centering
\includegraphics[scale=0.93]{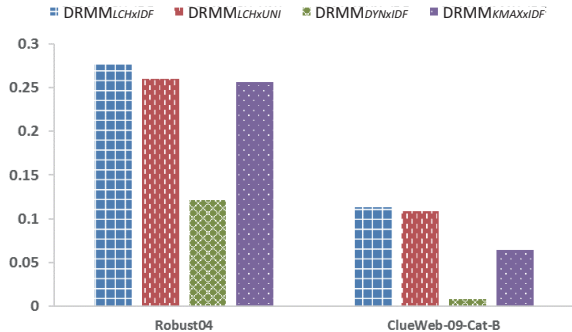}
\caption{Comparison of several simpler versions of DRMM over topic titles of the two test collections in terms of MAP.}\label{fig3}
\end{figure}

The comparison results over the topic titles on the two test collections in terms of MAP are depicted in Figure \ref{fig3}. As we can see, without the term gating network, DRMM$_{LCH\times UNI}$ performs slightly worse than the original DRMM. Specifically, the relative MAP drop of DRMM$_{LCH\times UNI}$ compared with DRMM$_{LCH\times IDF}$ is about $6.8\%$  and $3.5\%$ on Robust04 and ClueWeb-09-Cat-B, respectively. The results demonstrate the effectiveness of the differentiation of query term importance in relevance matching. Besides, we find that DRMM$_{DYN\times IDF}$ based on position-related signals performs significantly worse than the other two models based on strength-related signals (i.e., DRMM$_{LCH\times IDF}$ and DRMM$_{KMAX\times IDF}$). The results indicate that ad-hoc retrieval is more likely to be a strength-related task rather than a position-related task. When comparing DRMM$_{KMAX\times IDF}$ and the original DRMM$_{LCH\times IDF}$, we find that DRMM$_{KMAX\times IDF}$ works quite well on Robust04 but fails on ClueWeb-09-Cat-B. The possible reason is that the document length variation on Web data (i.e., ClueWeb-09-Cat-B) is much larger than that on news data (i.e., Robust04), leading to the failure of the $K$-max pooling method which has potential bias towards very long documents. This further demonstrates the effectiveness of our matching histogram mapping and the corresponding histogram based feed forward matching network.

\subsubsection{Impact of Term Embeddings}
Since we leverage a priori learned term embeddings in our model, we further study the effect of embedding dimensionality on the retrieval performance. Here we report the performance results on the Robust04 collection using term embeddings trained by CBOW model with $50$, $100$, $300$, and $500$ dimensions, respectively. As shown in Table \ref{tab:dimension}, the performance first increases and then slightly drops with the increase of dimensionality. Term embeddings of different dimensionality provide different granularity of semantic similarity; they may also require different amounts of training data. With lower dimensionality, the similarity between term embeddings might be coarse and hurt the relevance matching performance. However, with larger dimensionality, one may need more data to train reliable term embeddings. Our results suggest that $300$ dimensions is sufficient for learning term embeddings effective for relevance matching on the Robust04 collection.

\begin{table}
  \centering\small
  \caption{Performance comparison of DRMM over \mbox{different} dimensionality of term embeddings trained by CBOW on the Robust04 collection.}
  \begin{tabular}{c c c c c }
 \hline
    Topic & Embedding & MAP & NDCG@20 & P@20 \\\hline
    \multirow{4}{*}{Titles} & CBOW-50d & 0.268 & 0421 & 0.375  \\
    &CBOW-100d & 0.270 & 0.427 & 0.379 \\
    &CBOW-300d & 0.279 & 0.431 & 0.381 \\
    &CBOW-500d & 0.277 & 0.430 & 0.381 \\\hline
    \multirow{4}{*}{Descriptions} & CBOW-50d &  0.268 & 0.431 & 0.365  \\
    &CBOW-100d & 0.271 & 0.433 & 0.367  \\
    &CBOW-300d & 0.275 & 0.437 & 0.371 \\
    &CBOW-500d & 0.274 & 0.435 & 0.370 \\\hline
  \end{tabular}
  \label{tab:dimension}
\end{table}

\section{Related Work}
By formalizing ad-hoc retrieval as a text matching problem, deep matching models can be applied to this task so that features can be automatically acquired in an end-to-end way. In recent years, a variety of deep matching models have been proposed for the text matching problems. As mentioned before, we can categorize the existing deep matching models into two major types, namely representation-focused models and interaction-focused models. We have described several representative deep matching models in these two classes in previous sections including DSSM, C-DSSM, ARC-I, ARC-II and MatchPyramid. Here we will discuss some other related work in this direction.

In the class of representation-focused models, Qiu et al.~\cite{qiu2015convolutional} proposed Convolutional Neural Tensor Network (CNTN) for community-based question answering. The CNTN model is similar to ARC-I, using CNN to build the representations for each piece of texts. The major difference between CNTN and ARC-I is that CNTN employs a tensor layer rather than MLP on top of the two CNNs to compute the matching score between the two pieces of text. In \cite{socher2011dynamic}, Socher et al.~proposed an Unfolding Recursive Autoencoder (uRAE) for paraphrase identification. They first employed recursive autoencoders to build the hierarchical compositional text representations based on syntactic trees, and then conducted matching at different levels for the identification task. In \cite{yin2015multigrancnn}, Yin et al.~introduced MultiGranCNN which employs a CNN to obtain hierarchical representations of texts, and then computes the matching score based on the interactions between these multigranular representations.

In the class of interaction-focused models, Wang et al.~\cite{wang2015syntax} proposed Deep Match Tree (DeepMatch$_{tree}$) for the short text matching problem. Different from DeepMatch \cite{lu2013deep} which builds local interactions between texts based on semantic topics, DeepMatch$_{tree}$ defines interactions in the product space of dependency trees. A deep neural network is then leveraged for making a matching decision on the two short texts, on the basis of these local interactions. In \cite{wan2016match}, Wan et al.~introduced Match-SRNN to model the recursive matching structure in the local interactions so that long-distance dependency between the interactions can be captured. The proposed model was evaluated on two tasks, including community based question answering and paper citation \mbox{matching}.

Most of these deep matching models are designed for the semantic matching problem, which is significantly different from the relevance matching problem in ad-hoc retrieval. In this work, we introduce a model specifically designed for the relevance matching problem.

\section{Conclusions}
In this paper, we point out that there are significant differences between semantic matching for many NLP tasks and relevance matching for the ad-hoc retrieval task. Many existing deep matching models designed for the semantic matching problem thus may not fit the ad-hoc retrieval task. Based on this analysis, we propose a novel deep relevance matching model for ad-hoc retrieval, by explicitly addressing the three factors in relevance matching. The proposed model contains three major components, i.e., matching histogram mapping, a feed forward matching network, and a term gating network. Experimental results on two representative benchmark datasets show that our model can significantly outperform traditional retrieval models as well as state-of-the-art deep matching models.

For future work, we would like to leverage larger training data, e.g. click-through logs, to train deeper DRMM so that we can further explore the potential of the proposed model on ad-hoc retrieval. We may also include phrase embeddings so that phrases can be treated as a whole rather than separate terms. In this way, we expect the local interactions can better reflect the meaning by using the proper semantic units in language, leading to better retrieval performance.

\section{Acknowledgments}
This work was supported in part by the Center for Intelligent Information Retrieval, in part by the 973 Program of China under Grant No. 2014CB340401 and 2013CB329606, in part by the National Natural Science Foundation of China under Grant No. 61232010, 61472401, 61425016, and 61203298, and in part by the Youth Innovation Promotion Association CAS under Grant No. 20144310 and 2016102.

\bibliographystyle{abbrv}

\end{document}